\documentclass[conference,10p]{IEEEtran}

\usepackage[T1]{fontenc} 
\usepackage[utf8]{inputenc} 
\usepackage{microtype}
\usepackage{times}
\usepackage[scaled=0.85]{helvet}
\usepackage{graphicx}
\usepackage{ifthen}
\usepackage{xspace}
\usepackage{alltt}
\usepackage{latexsym}
\usepackage{url}            
\usepackage{amssymb}
\usepackage{amsfonts}
\usepackage{amsmath}
\usepackage{stmaryrd}
\usepackage{enumerate}
\usepackage{cite}
\usepackage{xspace}
\usepackage{paralist}
\usepackage{fancyvrb}
\usepackage{multirow}
\usepackage[table]{xcolor}
\usepackage{tabulary}

\usepackage{array}
\usepackage{booktabs}
\usepackage{textcomp}
\usepackage{nicefrac}

\newboolean{showcomments}
\setboolean{showcomments}{false}
\ifthenelse{\boolean{showcomments}}
  {\newcommand{\bnote}[2]{
	\fbox{\bfseries\sffamily\scriptsize#1}
    {\sf\small$\blacktriangleright$\textit{#2}$\blacktriangleleft$}
   }
   
  }
  {\newcommand{\bnote}[2]{}
   
  }

\graphicspath{{figures/}}

\newcommand{\commented}[1]{}

\newcommand{\eg}{\emph{e.g.,}\xspace}
\newcommand{\ie}{\emph{i.e.,}\xspace}
\newcommand{\etal}{\emph{et al.}\xspace}
\newcommand{\ct}[1]{`{\textsf{#1}'}\xspace}

\usepackage{url}            
\makeatletter
\def\url@leostyle{%
  \@ifundefined{selectfont}{\def\UrlFont{\sf}}{\def\UrlFont{\small\sffamily}}}
\makeatother
\makeatletter
\newcommand{\verbatimfont}[1]{\renewcommand{\verbatim@font}{\ttfamily#1}}
\makeatother
\verbatimfont{\scriptsize}

\usepackage{cleveref}

\newcommand{\approach}{\emph{EpiceaUntangler}\xspace}
\newcommand{\tool}{\emph{Epicea Task Clusterer}\xspace}
\newcommand{\dsOne}{\textsf{2devs}}
\newcommand{\dsTwo}{\textsf{devEval}}

\newcommand\martin{P1}
\newcommand\yuriy{P2}

\newcommand\tommaso{P3}
\newcommand\sebastian{P4}
\newcommand\anne{P5}
\newcommand\vincent{P6}
\newcommand\christophe{P7}
\newcommand\guillaume{P8}

\renewcommand{\quotation}[1]{``\emph{#1}''}

\begin{document}

\title{Untangling Fine-Grained Code Changes}


\author{
\IEEEauthorblockN{Mart\'in Dias$^1$, Alberto Bacchelli$^2$, Georgios Gousios$^3$, Damien Cassou$^1$, St\'{e}phane Ducasse$^1$}
\IEEEauthorblockA{\small 1: RMoD Inria Lille--Nord Europe, University of Lille --- CRIStAL, France \\ 
2: SORCERERS @ Software Engineering Research Group, Delft University of Technology, The Netherlands \\
3: Digital Security Group, Radboud Universiteit Nijmegen, The Netherlands
}
}

\maketitle

\begin{abstract}

After working for some time, developers commit their code changes to a
version control system. When doing so, they often
bundle unrelated changes (\eg bug fix and refactoring) in a single
commit, thus creating a so-called tangled commit. Sharing tangled commits
is problematic because it makes review, reversion, and integration of these commits harder
and historical analyses of the project less reliable.

Researchers have worked at untangling existing commits, \ie finding
which part of a commit relates to which task.
In this paper, we contribute to this line of work in two ways:
\begin{inparaenum}[(1)]
\item A publicly available dataset of untangled code changes, created
  with the help of two developers who accurately split their code
  changes into self contained tasks over a period of four months;
\item a novel approach, \approach,
to help developers share untangled commits (aka. atomic
  commits) by using fine-grained code change information.
\end{inparaenum}
\approach is based and tested on the publicly available dataset,
and further evaluated by deploying it to 7 developers,
who used it for 2 weeks. 
We recorded a median success rate of 91\% and average one of 75\%,
in automatically creating clusters of untangled fine-grained code changes.

\end{abstract}


\IEEEpeerreviewmaketitle


\section{Introduction} 

Version Control Systems (VCS), such as Git and Subversion, allow programmers
to control changes to source code and make it possible to find
who made each software change, when, and where.
This information is important to support both the coordination of developers
working in teams~\cite{Guzz2015a} and the creation of many recommendation 
and prediction systems related to software quality~\cite{Zimm2005a}.

Developers often bundle unrelated changes 
(\eg bug fix and refactoring) in a single commit~\cite{Herz11a}, 
thus creating a so-called \emph{tangled commit}, such as the following taken from Jaxen:\footnote{\url{http://jaxen.codehaus.org}, commit: svn-1252, 2006-11-09} 

\begin{Verbatim}[fontsize=\scriptsize]
-----------------------------------------------------
r1252 | elharo | 2006-11-09 [...] | 2 lines

Pulling getOperator up into BinaryExpr per Jaxen-169

[...]
Index: src/java/main/org/jaxen/expr/AdditiveExpr.java
=====================================================
--- src/[...]/AdditiveExpr.java	(revision 1251)
+++ src/[...]/AdditiveExpr.java	(revision 1252)
@@ -61,7 +61,7 @@
  * 
- */public interface AdditiveExpr extends BinaryExpr
+ */
+public interface AdditiveExpr extends BinaryExpr
 {
-    String getOperator();
 }
\end{Verbatim}

The tangled commit above
contains both a refactoring (the move of \ct{getOperator} to a 
different place, not shown in this extract), and code formatting 
(the move of an interface definition to its own line).
Sharing tangled commits is problematic as they make 
code review, reversion, and integration harder and historical analyses of the
project less reliable~\cite{Herz13a}. For example, even integrating the 
code formatting change included in the aforementioned 
commit (without the refactoring) would be a demanding task.

Untangling existing commits (\ie finding how to separate parts of 
a commit relating to different tasks) is an open research problem.
Herzig and Zeller presented the earliest and most significant results 
in this area~\cite{Herz11a}: They implemented the first algorithm that can automatically untangle commits given artificially tangled ones.

In this paper, we expand on this previous work by:
\begin{inparaenum}[(1)]
\item working in an untyped setting where a part of the approach
by Herzig and Zeller is inapplicable;
\item considering fine-grained code change information gathered during
  development (\eg time at which each line has
  changed and all versions of each line); and
\item evaluating the resulting approach both on
data generated by programmers who manually label it and
with programmers working on real-world development tasks.
\end{inparaenum}

The ultimate goal of our work is to help developers of
dynamically-typed code share untangled commits. To that end,
we:
\begin{inparaenum}[(1)]
\item asked 7 developers to
  manually cluster changes for each of their commits using a dedicated tool,
  for a period of 4 months;
\item manually validated the generated data, selecting the data recorded 
by two of these developers, and computed
a number of features based on their fine-grained code changes;
\item modeled the problem of predicting whether two fine-grained changes
  belong together, with a variety of machine learning approaches,
determined the most appropriate one, and identified the most
significant features;
\item designed an algorithm that uses the machine learning result to 
  propose an automatic clustering of any
  tangled commit and developed a corresponding tool, \approach; and 
\item evaluated the effectiveness of our approach with developers
who used \approach in their daily work for two weeks.
\end{inparaenum}

Our results show that three features are especially important to
perform clustering of fine-grained code changes: (1) the time between two
changes; (2) the number of other changes between two changes; and (3) whether
the two changes modify the same class. By modeling these features with
Random Forests~\cite{Brei2001a}, we identify whether two changes belong to
the same commit with an accuracy of 95\%, if training and testing
on the same developer, and more than 88\% if tested on a different developer.
A set of 200 manually clustered fine-grained code changes
(\ie the equivalent of a few days of work) was sufficient
to reach good performance.
When deploying \approach with new developers during their daily tasks,
we recorded an average success rate of 75\% and a median one of 91\%.





\section{Problem Description} \label{sec:problem}

When developers want to share their work in a VCS, they will, more often
than not, realize that they have done more than one activity, \eg fixed a
bug, reformatted a method, and fixed a typo in a comment.
Sharing everything in a single \emph{tangled commit} is regarded as
bad practice because it makes the following activities more difficult:

\begin{itemize}
\item \emph{Review} -- Reviewers have to understand the code changes 
  of all the activities \emph{at once}~\cite{Tao2012a, Bacc2013a, GZSD14};
\item \emph{Reversion} -- Developers have to revert all changes of a
  problematic commit even when only the code change of one activity is problematic~\cite{Guzz2015a};
\item \emph{Integration} -- Integrators have to merge or reject whole
  commits, \eg they will typically reject a code formatting operation and
  a bug fix included in the same commit~\cite{Uqui12b};
\item \emph{Historical analysis} -- Researchers need to associate
  activities to files to conduct statistical analyses while,
  \eg mining software repositories~\cite{Herz13a}.
\end{itemize}

\subsection{Existing Solutions for Tangled Changes} \label{sec:exSolution}

To avoid tangled commits, developers could organize their work so that,
at commit time, only one activity's code is to be shared. This
requires frequent commits and interruptions in the developer's work
flow\cite{Apache2003SVNBestPractice,Beck00a,Fowl99b,Stein12b}. Even
with a lot of discipline, there will be times when a developer will
have to split changed code into several commits.

To separate code from several activities into different commits, some
tools (\eg \ct{git add}) let the user selects which files and lines to
commit first. Being line based, these tools share the following
problems:
\begin{inparaenum}[(1)]
\item The code present at commit time might be
  \emph{incomplete}\cite{Nega12a}: Each change to a line shadows
  previous changes of the same line, thus making it impossible to commit the
  line as it was before the last change;
\item a commit resulting from a manual selection of a subset of all
  changed lines might be \emph{invalid}: \eg a developer
  might commit the beginning of a function definition but not the end; and
\item changed lines are shown in the order they appear in their
  files irrespective of their modification time: This makes it difficult
  for developers to select lines changed closely in time.
\end{inparaenum}

A great source of inspiration for us comes from Herzig \etal
\cite{Herz11a,Herz13a}, who implemented an algorithm to
automatically untangle commits. Their
algorithm uses several \emph{confidence voters} to decide whether two lines
of a tangled commit should be put in the same cluster. 
They aggregate the results of each confidence voter
into a single score, and then use the concepts 
of a multilevel graph-partitioning algorithm by Karypis and Kumar~\cite{Kary1995a}
to generate the clusters. Their voters include:

\begin{itemize}
\item \ct{FileDistance}: the number of lines between the two lines if they
  are both in the same file;
\item \ct{PackageDistance}: the number of different package name segments within the package names of the changed files;
\item \ct{CallGraph}: the difference between the call graphs of the program
  with each line change applied separately;
\item \ct{ChangeCouplings}: the frequency with which the files both lines
  were changed into are committed together, using the work from
  Zimmermann \etal \cite{Zimm2005a};
\item \ct{DataDependency}: a boolean indicating if the two lines read or
  write the same variable(s).
\end{itemize}

In the work by Herzig \etal we see the following limitations:

\begin{itemize}
\item[\emph{Dependence on static-analysis}:] The voters \ct{CallGraph} and
  \ct{DataDependency} rely on static analyses that might not be
  possible for dynamically-typed programming languages, or that might
  be available in a weaker form;
\item[\emph{Incompleteness}:] The tangled commits used as input to the
  algorithm suffer from the incompleteness problem described earlier
  in this section: If a line is changed twice before a commit, the
  commit only contains the latest version of the line, shadowing a
  previous version of the line which could have been part of an
  untangled commit;
\item[\emph{Artificiality}:] The validation by Herzig \etal relies on a
  classification of 7,000 existing commits done by the researchers without
  feedback from each project's experts. We believe that only the
  author of each commit can, at commit time, best organize his changes
  into untangled commits. Moreover, the untangling algorithm by Herzig \etal 
  relies on the knowledge of the expected number of untangled commits
  for a particular tangled one. With the goal of helping developers
  creating untangled commits, we do not have access to this information.
\end{itemize}

\subsection{Addressing the Current Limitations}

In our work, we propose to alleviate the aforementioned limitations by
\begin{inparaenum}[(a)]
\item expanding the setting to a dynamically-typed environment where some
  kinds of analyses are not available;
\item using fine-grained code changes that we collect during development
  sessions;
\item relying on developer-approved data for the validation of untangling approaches.
\end{inparaenum}
This results in the following requirements for the approach, \approach, that we present in this paper:

\emph{The Dynamically-Typed Setting}: Whereas the approach of Herzig \etal
relies on static analysis of Java programs to untangle commits,
our approach helps developers to create untangled commits in an environment
that is dynamically-typed. Certain types of static analysis, \eg
accurate call graph analysis, is not possible for dynamically-typed
languages. Therefore, our approach cannot rely on such static analysis.

\emph{Fine-Grained Changes}: In modern integrated development environments (IDEs), 
tools can be notified each time a software artifact is changed and saved.
As a result, a tool could listen to all
fine-grained changes made by developers and, at commit time,
present the developer a list of all the changes they have done. For
example, a developer changing and saving the source code of a method
3 times will result in 3 fine-grained changes. This is in contrast with most tools
that only present the latest version of each changed line;
this requirement tackles the \emph{incompleteness} limitation.

\emph{Developer-Approved Data}: The untangling algorithm should
be based on data created by developers who personally untangle the
tangled commits that they produced in the first place. The final version of the approach
should provide each developer, at commit time, with a list of
the automatically untangled commits containing their fine-grained
changes: Each developer could then reorganize these
automatically-computed clusters of changes. Results must be validated by
comparing the change clusters that are automatically computed against
the reorganization done by the developer in a manual way.




\section{Proposed Solution} 



In a nutshell, our solution is to develop an approach
and associated tools to help developers share untangled commits.
The tools log all the fine-grained changes made by developers
as they change the source code. When a developer wants to commit her changes, 
the tool, based on an analysis of the recorded information, 
presents several automatically-computed clusters of changes: 
Each cluster represents a distinct activity of the developer since last commit. 
The developer may then add a comment to each cluster and, if necessary,
adapt the automatic clustering (by adding/removing clusters and moving
changes to different clusters). Once the developer validates the clusters,
the tool generates one commit per cluster and publishes them to the repository.
In the following section, we present our solution decomposed in individual
parts.

\subsection{Epicea: Event Modeler with Fine-Grained Changes}

\begin{figure}
  \begin{center}
    \includegraphics[width=0.86\columnwidth]{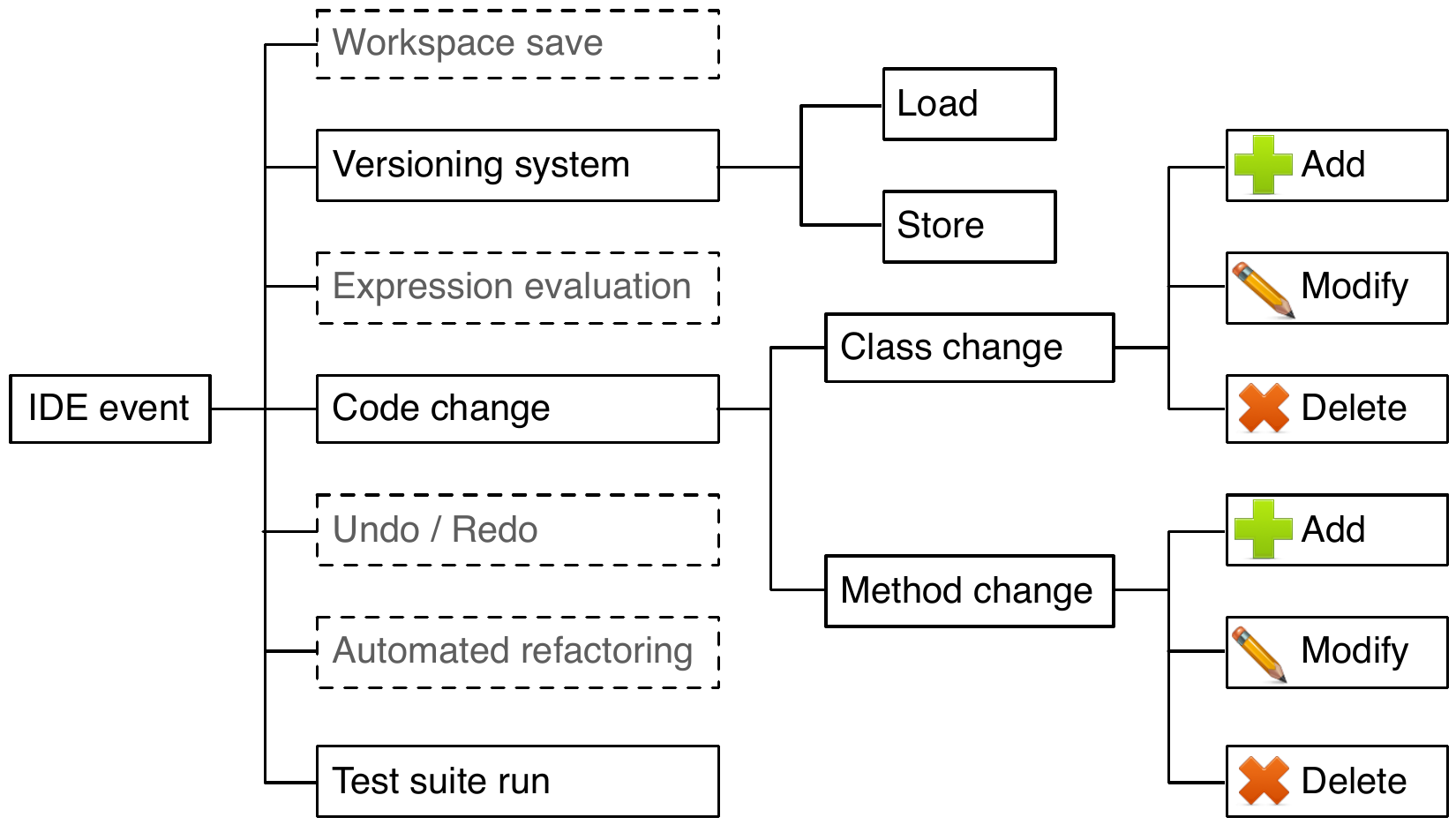}
  \end{center}
  \caption{IDE events recorded by Epicea; dashed ones are not considered.}
  \label{ide_events}
\end{figure}

Central to our approach is the collection of fine-grained information.
To conduct this task, we use Epicea~\cite{Dias13a}, a tool we developed to model IDE events. In essence, Epicea listens to actions taking place in the IDE and records different types of events. A simplified version of the events recorded by Epicea is shown in \Cref{ide_events}. Epicea records complete information of these events (\eg whether a test run failed), including a timestamp. As previous studies (both in Eclipse~\cite{Hatt2010b} and in Smalltalk~\cite{Robb2007a}) showed that save-based recording produces reliable fine-grained code change data, we record code change operations (add, modify, and delete classes and methods) every time the user saves the code. Epicea is invisible to the user as there is no impact on performance. Epicea stores the collected data as a sequence of serialized objects in plain text files.

\subsection{Voters} \label{sec:voters}

Once the data is collected, we have to characterize it in a way that 
it can be used for generating untangled changes.
Similarly to Herzig \etal~\cite{Herz11a,Herz13a}, as a first step
we model our clustering task as a binary classification problem:
For all the potential pairs of recorded fine-grained changes,
we want to determine whether they belong in the same cluster.
To this end, we implement a number of features or, maintaining
the term used by Herzig \etal~\cite{Herz11a}, \emph{voters},
which describe different relations between the considered changes.
Our voters (detailed in \Cref{tab:voters}) span the following six dimensions:

\begin{enumerate}

\item \emph{Code structure}: Although dynamic languages make it difficult
to conduct static analysis, it is possible to compute basic relations.
Our three voters in this dimension consider whether two changes
happen in the same package, class, and/or method.

\item \emph{Content}: This voter returns true if the two changes to a method are only 
source-code reformats, \ie if the \emph{abstract syntax tree} of a method remains 
the same after a change on it. This voter should help linking changes regarding 
refactoring actions.

\item \emph{Testing}: Epicea records test runs. The rationale of this voter
is that two changes happening between runs of the same test could be related
to the same task (\eg this should hold in the case of test-driven development).

\item \emph{Spread}: These voters measure the distance between the two
considered changes, considering time passed and number of other changes
in between. We expect close changes to be more related.

\item \emph{Message sending}: This dimension analyzes whether the changes
involve related message sending (also known as `method invocations', in languages such as
Java or C\#).

\item \emph{Variable accessing}: This dimension computes relations between the
variables accessed by the two changes: For example, a change that adds a new instance variable
to a class may be related to a change that adds an usage of the same variable in a method.

\end{enumerate}

The input of each voter is a pair of changes, and the output is of the type specified in
column `Type' of \Cref{tab:voters}.




\begin{table*}
  \centering
\caption{Different voters tested in our investigation.}
  \begin{tabulary}{\linewidth}{l|l|l|L}
  
  {\bf Voter Name} & {\bf Dimension} & {\bf Type} & {\bf Relation between the two considered changes}\\


  \midrule


  \rowcolor[gray]{0.97}samePackage &  Code structure & Boolean & They involve the same package.\\

  sameClass &  Code structure & Boolean & They involve the same class.\\

 \rowcolor[gray]{0.97}sameSelector & Code structure & Boolean &  They involte a method with the same name (regardless its class).\\ \midrule

  bothCosmeticChanges & Content & Boolean &  They are both cosmetic (\ie pretty-printing---both versions of the method return the same result).\\ \midrule
  
  sameTestRun &  Testing & Boolean & They are modified between the same unit-test runs.\\ \midrule

  \rowcolor[gray]{0.97}numberOfEntriesDistance &  Spread & Numeric & How close they are in the history; the voter computes number of other changes between them.\\

 timeDifference &  Spread & Numeric & How close in time they are in the history; the voter computes the seconds between them. \\ \midrule

  \rowcolor[gray]{0.97}reciprocalMessageSends & Message sending & Nominal & They invoke each other; it computes 0, 1 or 2 if, respectively, no, one, or both call the other.\\ 
  
  numberOfSharedMessageSends & Message sending & Numeric & They share a number of the same message sends. \\

 \rowcolor[gray]{0.97}numberOfSharedMessageSendsInDelta & Message sending & Numeric & They add or remove a number of the same message sends.\\ \midrule

  numberOfVariableAccesses & Variable accessing & Numeric & One change modifies or adds definitions of instance variables, the other accesses some of them. \\ 
  
 \rowcolor[gray]{0.97}numberOfSharedVariableAccesses & Variable accessing & Numeric & They access a number of the same instance variable names.\\

  numberOfSharedVariableAccessesInDelta & Variable accessing & Numeric & They start or stop accessing a number of the same variable names.\\

  \bottomrule
\end{tabulary}
\label{tab:voters}
\end{table*}


\subsection{Machine Learning Approaches} \label{sec:ml}
Our approach computes the values for each voter for each pair of changes
(for performance reasons, we only consider fine-grained change pairs that are
less than 3 days apart); to aggregate these values and train models that 
would predict whether two changes should be in the same cluster,
we use machine learning (ML).

We consider three well-known machine learning algorithms
that can handle binary classification~\cite{Hast2001a}:
\begin{inparaenum}[(1)]
\item binary logistic regression (\ct{binlogreg}), 
\item na\"{i}ve bayes (\ct{naivebayes}), and 
\item random forests~\cite{Brei2001a} (\ct{ranforest}).
\end{inparaenum}
We chose these algorithms not only because they have been applied
successfully to a number of data mining tasks related to software engineering,
but also because they make quite different assumptions on the underlying data
and model (\eg \ct{naivebayes}relies on the conditional independence
assumption, \ie the value of a voter is unrelated to the value of the others,
and \ct{binlogreg} requires each observation to be independent
and linearity of independent variables and log odds), thus they can
offer different interpretations.
The choice of the most appropriate machine learning algorithm
is based on the empirical data collected during the experiment.

This machine learning step takes as input the values computed by the voters 
for two particular changes, and it outputs the probability of the two changes 
belonging to the same cluster.

\subsection{Clustering} \label{sec:clustering}

The last necessary step in our approach is to take the output of the machine
learning step, computed on each pair of changes, and aggregate it to form
the clusters of changes for the user.

In this method, each change is initially considered to be a cluster of its own. Then pairs of clusters are successively selected by their maximum scores and merged. The result of this method is a \emph{dendrogram}, which is a binary tree that represents the nested clustering of code changes. In this dendrogram, each non-leaf node has a \emph{similarity level} that represents how similar are both children. In our problem, a similarity level of 1 corresponds to two clusters that must be merged, while a level of 0 corresponds to the opposite decision. 

Finally, the desired clustering of code changes is obtained by cutting the dendrogram at some \emph{similarity threshold}. Using a too low threshold produces too many small clusters, while a threshold that is too high produces a single cluster. The choice of the most appropriate similarity threshold depends on the change set and, similarly to the machine learning approach, is based on the empirical data collected during the experiment.

The output of this step is the set of independent clusters of fine-grained changes, which is eventually displayed to the user with a dedicated user interface.




\section{Research Method} 

In this section, we describe how we structure our research in terms of research questions, we present the research settings, and we outline our research method. 

\subsection{Research Questions}

The ultimate goal of our work is to help developers of
dynamically-typed code share untangled commits.
For this we devise and test the approach we previously described to untangle code changes at a fine level of granularity. Accordingly, we structure our empirical investigation through the following three research questions:

\begin{enumerate}[RQ1:]
  \item {\bf Which voters are significant to untangle fine-grained code changes?}\\ 
  With this question we aim to understand which are the most important voters in our untyped setting. To answer this research question, we consider the
machine learning task of deciding whether two changes should belong to the same cluster. In doing so, we also determine which machine learning approach among the three we test, is better suited to model the problem through our voters.

  \item {\bf How effective is a machine learning model based on the significant voters in untangling historical fine-grained code changes?}\\
  Once we find the most significant voters and the best machine learning approach, we are interested to know their performance in predicting whether two changes should belong to the same cluster. We also want to investigate the effect asserted by individual developers' working styles on prediction performance;
 for this we train and test the machine learner on data generated by different developers (\eg training on one developer's data and testing on another developer's data).
  
  \item {\bf How effective is a tool based on the best voters and machine learning approach, when deployed with developers working on their daily tasks?}\\
  Finally, we want to devise an approach \approach, based on the best machine learner and voters, to generate clusters and present them with a graphical user interface. We want to test its effectiveness when deployed with participants  
  \begin{inparaenum}[(1)]
  \item whose data should not have been used for training the classifier, and 
  \item who should be working on their usual development tasks.
  \end{inparaenum}
  
\end{enumerate}

\subsection{Research Settings}

Our study took place with professional developers, researchers, and students using the Pharo environment.\footnote{Pharo: \url{http://pharo.org/}} Pharo is an open-source dialect of Smalltalk and implementation of its programming environment. Pharo was forked from Squeak\footnote{Squeak: \url{http://www.squeak.org/}} in 2008 and it is rapidly evolving. Currently, Pharo has around 60 worldwide contributors, it is used by more than 15 universities to teach programming and by 10 research groups to build tools, and more than 50 companies are using it in production.

We chose Pharo as a case study for two main reasons: \begin{inparaenum}[(1)]
\item the Pharo open-source community of developers has been receptive, since its inception, to welcome and thoroughly evaluate research tools (\eg~\cite{Reng10b,Hora13a,Hora14a}); and
\item the programming language, the development environment, and the versioning system are tightly integrated.
\end{inparaenum}
The later feature allows for a fast prototyping of an approach to record fine-grained code changes and interaction with testing and the versioning system. The former feature allow us to collect fine-grained data about code changes and IDE interactions from participants doing real-world development work. It also enabled us to deploy our resulting tool with more participants to evaluate its results. Moreover, many research tools tested within Pharo later became integral part of the environment (\eg~\cite{Verw11a}); we want both to improve the state of the art in untangling code changes and to create an approach that can be used in real-world scenarios. 

\subsection{Research Steps}


\subsubsection{Fine-grained data generation and collection} 
\begin{figure}
  \begin{center}
    \includegraphics[width=0.96\columnwidth]{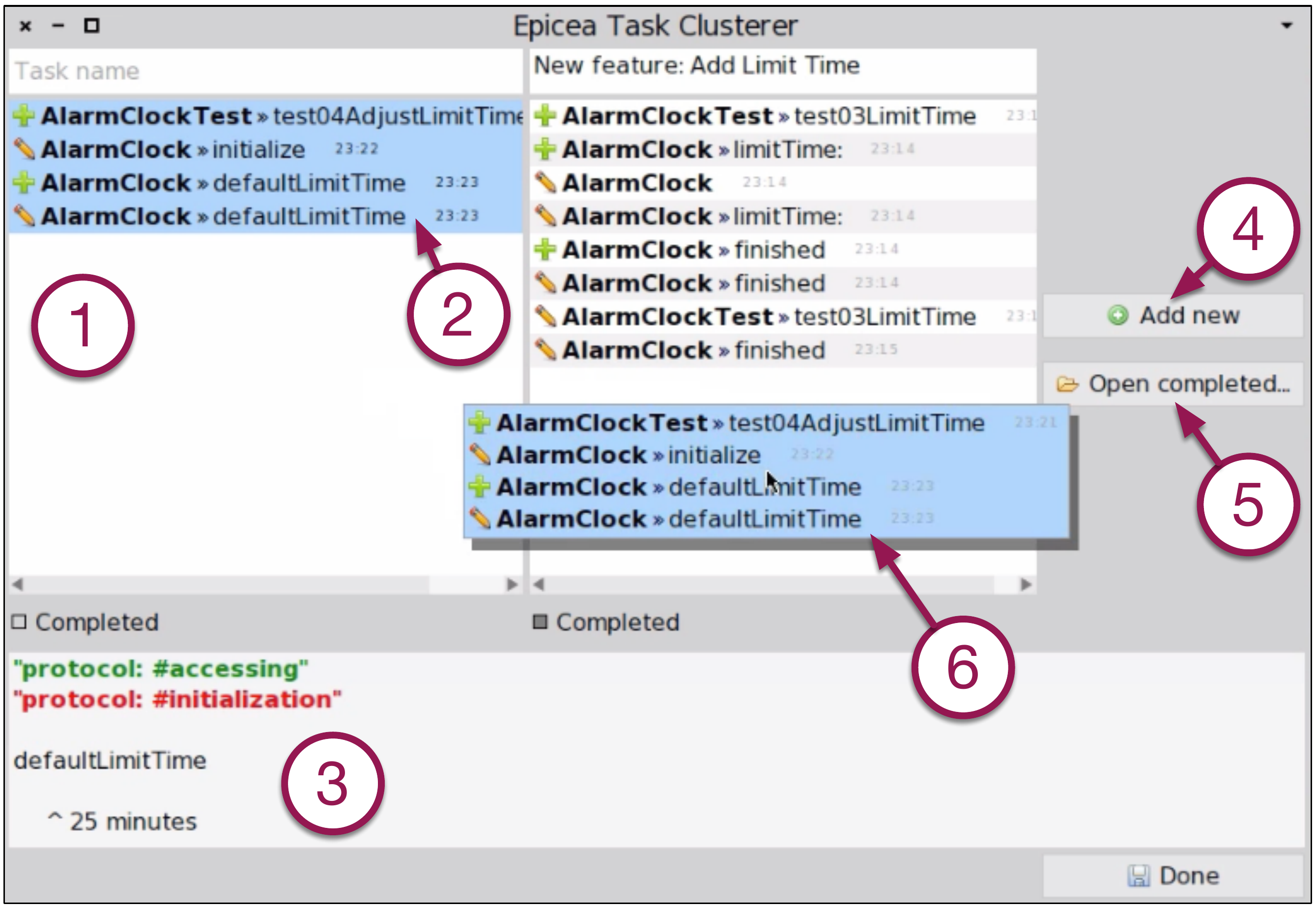}
  \end{center}
  \caption{UI used in training stage. The user manually clusters the changes.}
  \label{screenshot-training}
\end{figure}

To answer our first two research questions, we need a \emph{ground truth} to train and test our voters and machine learning approaches. Such a ground truth should be a reliable dataset containing fine-grained code changes correctly split into tasks by their authors. To obtain this, we contacted 7 participants actively contributing to Pharo, including the first author of this paper. We asked them to install Epicea and to use the tool (\ie \tool (ETC), \Cref{screenshot-training}) that we devised to manually cluster their fine-grained code changes. We showed a screencast\footnote{Available at: \url{https://www.youtube.com/watch?v=fQVWuMQUBew}} demoing the tool to all the participants before they started using it, so that they could understand the goal of the experiment and adapt their workflow accordingly. Every time the participants decided to commit their code to the versioning system, during their normal work, the ETC's interface would appear (as in \Cref{screenshot-training}) with a list of all the fine-grained changes, since the previous commit, that the user had to manually cluster into tasks.


\begin{table}[ht]
\begin{center}
\caption{Participants' information}
\label{tab:participants}
 \begin{tabular}{l | c | rr | r}
 \multirow{2}{*}{\bf P\_ID} & {\bf current} & \multicolumn{3}{c}{\bf programming experience (in months)} \\
  & {\bf role} & {\bf overall} & {\bf industrial} & {\bf with Pharo} \\
  \midrule 
\multicolumn{5}{c}{\bf }\\  
\multicolumn{5}{c}{\bf Data generation and collection phase}\\
\martin & Ph.D. student & 168 & 60 & 48 \\
\rowcolor[gray]{0.97} \yuriy & Ph.D. student & 48 & 36 & 24 \\   \midrule 
\multicolumn{5}{c}{\bf }\\  
\multicolumn{5}{c}{\bf Evaluation in real-world development phase }\\
\tommaso & Ph.D. student & 180 & 18 & 36 \\
\rowcolor[gray]{0.97} \sebastian & software engineer & 132 & 72 & 13 \\
\anne & associate professor & 72 & 12 & 24 \\
\rowcolor[gray]{0.97} \vincent & Ph.D. student & 72 & 11 & 11 \\
\christophe & software engineer & 180 & 10 & 30 \\
\rowcolor[gray]{0.97} \guillaume & software engineer  & 60 & 18 & 36 \\ \midrule
\end{tabular}
\end{center}
\end{table}


In detail, the main user interface of \tool (shown in \Cref{screenshot-training}) works as follows: In the top pane, each column (\eg Point 1) represents a task (to group an activity of the user), and each item in a column represents a code change (\eg Point 2). Each code change is in a \emph{ClassName}>>\emph{methodName} format, and the icon shows the type of change (as in \Cref{ide_events}). The bottom pane (Point 3) shows the details of the selected change, in a \emph{unified-diff} format. The user can review the listed changes and perform three actions to specify the expected clustering for them: Add a new empty task/cluster (Point 4), reopen an already closed task/cluster (Point 5), and move changes between columns (with drag and drop, Point 6). Once the clustering task is completed, the user presses the button \ct{Done}, and the interface disappears.

\subsubsection{Data analysis and evaluation of voters} Once the participants concluded the data collection period of 4 months, we conducted exploratory data analysis~\cite{Onei2013a} on the generated clustered changes. The data generated by five users was extremely sparse and inconsistent; these users confirmed this explaining that they could not afford the time required by \tool to review each change made during the experiment period. We removed this data and kept the data generated from the remaining two users (including the first author of this paper) whose features are described in the top half of \Cref{tab:participants}. \Cref{tab:dsone} describes the resulting dataset (\dsOne).


\begin{table}[ht]
\begin{center}
\caption{Descriptive statistics of dataset \dsOne}
\label{tab:dsone}
\begin{tabular}{l | rr | rrrr }
\multirow{2}{*}{\bf P\_ID} & \multicolumn{2}{c}{\bf Total number of }& \multicolumn{4}{c}{\bf Changes per cluster}\\
 & {\bf changes} & {\bf clusters} & {\bf Mean} & {\bf Median} & {\bf St. Dev.} & {\bf Max} \\ \midrule
 
 \martin & 15,175 & 298 & 50.9 & 8 & 153.1 & 1,582\\
\rowcolor[gray]{0.97}  \yuriy & 9,601 & 119 & 80.7 & 16 & 151.9 & 812

\end{tabular}
\end{center}
\end{table}


Using \dsOne~we answered RQ1 and RQ2. 
As previously detailed (\Cref{sec:ml}), we used machine learning to 
identify pairs of changes belonging to
the same commit, by modeling it as a binary classification. 
For all potential pairs of changes in \textsf{2devs}, we
calculated values for all the voters in \Cref{tab:voters} and labeled
with `\textsf{true}' if the changes belonged to the same commit or
`\textsf{false}' otherwise. As our dataset was unbalanced (the \textsf{false} class 
overruled the \textsf{true} one by a ratio of 4:1), we adjusted to avoid overfitting. 
Models where thus trained with a 
ratio of 2:1 samples for the \textsf{false} and \textsf{true} class respectively.

\textbf{Evaluation of voters.} To evaluate each trained model, we used standard machine learning metrics~\cite{Hast2001a}, 
such as precision (\textsf{prec}), recall (\textsf{rec}), accuracy (\textsf{acc}),
the Area Under the receiver operating characteristic Curve ({\textsf{auc}}) and
the F-measure ({\textsf{f.measure}}).  Models where trained with an increasing
number of samples as input ($10^4$ to $10^6$ samples) to determine the
minimum number of samples required to obtain adequate performance. At each input
size, we used random selection 10-fold cross validation to evaluate model
stability and reported results based on the mean of the 10 runs.
We selected the best classifier and applied a classifier-specific process to
rank voters according to their importance in the classification. Then we
incrementally trimmed the voter set starting from the least important feature
until the performance of the classifier was severely impacted. Finally, we
retrained the best classifier with the trimmed voter set and used that as our final
prediction model. The final model was then exposed as a web service that
\approach used to drive the change untangling process to answer
RQ3.

\subsubsection{Deployment and evaluation with developers} Once we 
completed the creation and evaluation of the best ML approach and features 
on dataset \dsOne, and obtained promising results, we created
the corresponding implementation in \approach, a tool that developers
can use in real-world development.

\begin{figure}
  \begin{center}
    \includegraphics[width=0.7\columnwidth]{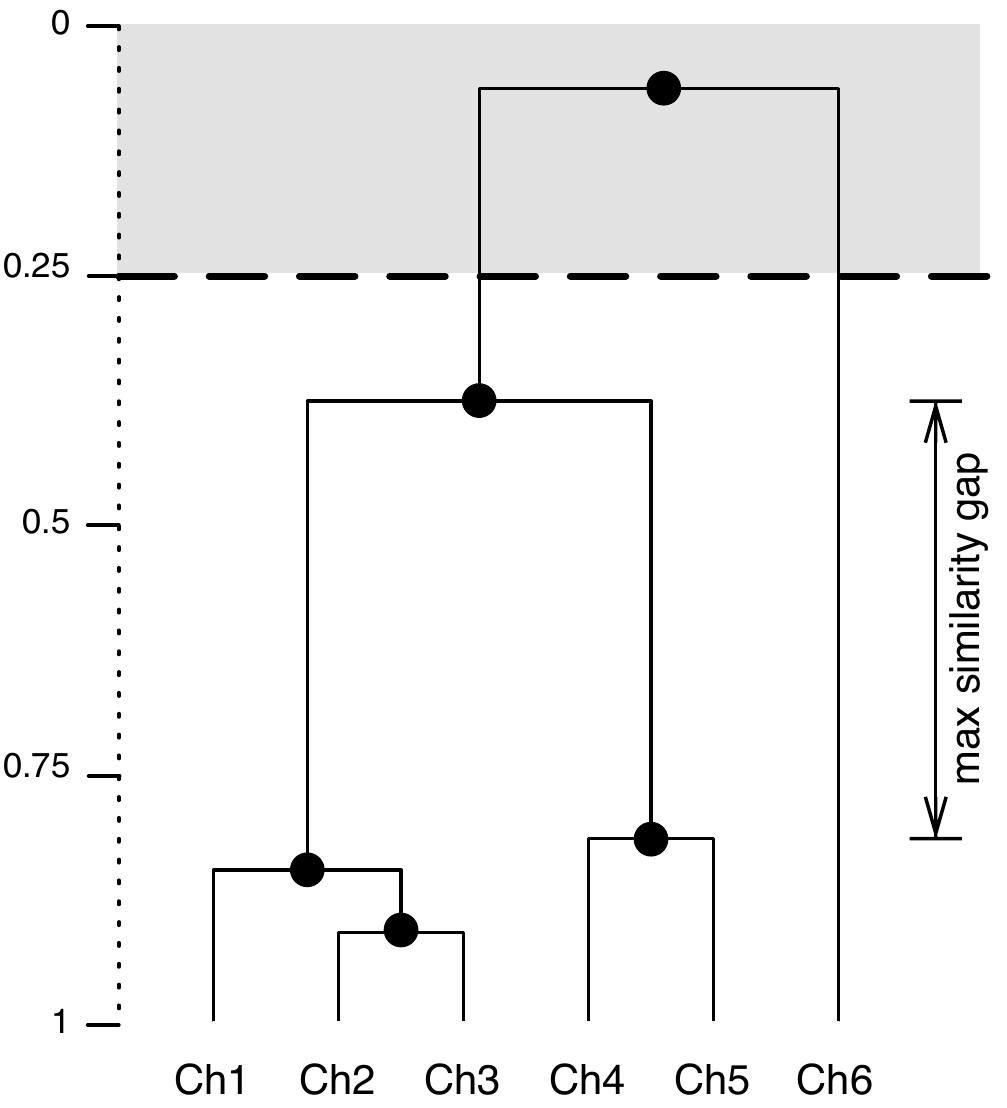}
  \end{center}
  \caption{Determining the \emph{similarity threshold} to cut the dendrogram.}
  \label{fig:dendrogram}
\end{figure}

During developer's work, \approach records the fine-grained change information, exactly as done for
the data collection phase. When the developer wants to commit,
our approach computes the values for all the significant voters for each pair
of code changes, and queries the web service implementing the
final model of the ML classifier. For each pair, the
web service returns a score between 0 and 1, indicating the probability 
that the two changes belong to the same cluster, according to the trained model.
 \approach aggregates all the scores to 
form clusters using agglomerative hierarchical clustering method (see \Cref{sec:clustering}).
This method outputs a dendrogram, which has to be cut at some similarity threshold to obtain the clusters of changes. We created a testbed with \ct{change set-expected clustering} pairs whose purpose is to help us to conceive a good function for obtaining the similarity threshold for cutting the dendrogram. In \Cref{fig:dendrogram} we illustrate the function. The similarity threshold we chose corresponds to the maximum similarity gap between all nodes whose similarity level is less than 0.25. The intuition behind taking the maximum similarity gap is that continue merging code changes together is not worth, because the meaningful clusters have already been detected. The reason to use 0.25 as a lower bound is that we observed from data that such a low likelihood indicates in most cases changes that should not be merged.

This process happens in the background: After the developer decides to commit,
she sees an interface similar to that used to generate the data for \dsOne~
(\Cref{screenshot-training}), with the difference that the clusters
are already pre-computed by the tool. Then, the user browses the clusters and reorganizes the changes in case the pre-computed clusters are wrong.

\textbf{Evaluation of clustering.} To conduct this evaluation, we recruited six participants, whose
features are described in the bottom half of \Cref{tab:participants}.
They all used \approach for 2 weeks. To evaluate the clustering, each participant
was asked to confirm whether the automatic clustering was correct; if not 
they could rearrange changes to the correct clusters.
We used the resulting data to evaluate the accuracy of our approach.

To measure the success rate of our approach, 
\ie how similar the \emph{computed clustering} (from our
algorithm) is to the \emph{expected clustering} (from the developer), 
we need to know the ratio between the number of successfully
clustered changes and the total number of changes. To know if a change
has been successfully clustered, we must find which computed cluster best
matches which expected cluster. 

\begin{figure}
  \begin{tabular}{cc}
  
  \begin{minipage}[c]{0.35\columnwidth}
    \includegraphics[width=\textwidth]{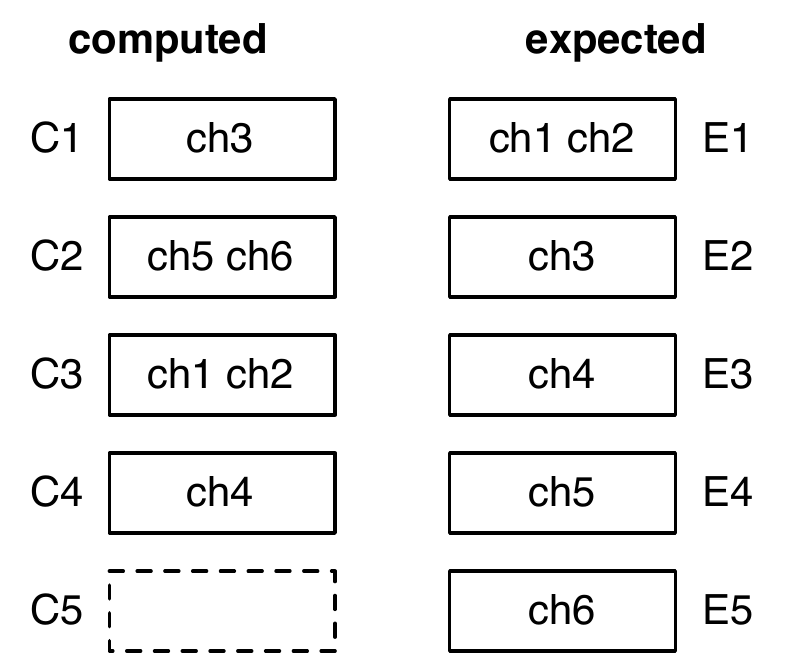}
  \end{minipage}
 
  &
 
  \begin{minipage}[c]{0.48\columnwidth}
    \begin{tabular}{r|ccccc}
      & \textbf{E1} & \textbf{E2} & \textbf{E3} & \textbf{E4} & \textbf{E5} \\
      \toprule
\textbf{C1} & 0 & 1 & 0 & 0 & 0 \\
\textbf{C2} & 0 & 0 & 0 & \textonehalf & \textonehalf \\
\textbf{C3} & 1 & 0 & 0 & 0 & 0 \\
\textbf{C4} & 0 & 0 & 1 & 0 & 0 \\
\textbf{C5} & 0 & 0 & 0 & 0 & 0 \\
    \end{tabular}
  \end{minipage}

\end{tabular}

\caption{Comparison between a computed clustering and an expected
  clustering. On the left-hand side, each box represents a cluster of
  changes. The computed clustering contains 4 clusters labeled from $C1$ to
  $C4$ (cluster $C5$ is a \emph{virtual cluster} to ease comparison). 
  The expected clustering has 5 clusters: $E1$ to $E5$. On
  the right-hand side, the matrix shows the corresponding Jaccard indexes.}
\label{fig:computed-vs-expected}
\end{figure}

Figure \ref{fig:computed-vs-expected} shows a sample comparison
between a computed clustering and an expected clustering. The matrix on
the right represents the \emph{Jaccard indexes} computed for each pair
of clusters; this index is defined as using the following formula:

{\small $$ J_{CiEj} = \frac{ \left | Ci \cap Ej \right | }{ \left | Ci \cup  Ej \right | } $$}

This Jaccard index represents how much two sets coincide.
It ranges from 0 to 1, where $1$ means the two sets are equal (\eg $C3$ and $E1$ in Figure
\ref{fig:computed-vs-expected}) and $0$ means the
two sets have nothing in common (\eg $C4$ and $E2$ in \Cref{fig:computed-vs-expected}).

From the resulting matrix we want to know which computed cluster matches
which expected cluster. This can be obtained by maximizing the sum of
the Jaccard indexes over all permutations. For the sample in
\Cref{fig:computed-vs-expected}
%
%
%
the maximum sum over all the permutations ($3.5$) is attained for this
set of pairs: 

{\small  $$ Matching = \{ (C1,E2) (C2,E4) (C3,E1) (C4,E3) (C5,E5) \} $$} 
  
We compute the success rate of our algorithm using the following formula:

{\small $$ SuccessRate = \frac{\#SuccessfullyClusteredChanges}{\#Changes} $$}

A change $ch_i$ is \emph{successfully clustered} if the
computed and expected clusters that contain $ch_i$ are in the same pair
of the $Matching$ set. In \Cref{fig:computed-vs-expected},
all changes are successfully clustered except $ch6$.
This gives us a success rate of $\nicefrac{5}{6}=0.83$.




\section{Results} 


In this section we answer our research questions, by describing the results
we obtained in our evaluations.

\subsection{What Are the Dominant and Significant Voters?}

As a first step to answer our first research question, we use all the
machine learning approaches we consider on the collected data
and we evaluate whether an approach performs undoubtedly better.
\Cref{tab:classifierperf} reports the results 
of the classification performance of each machine learning approach
for predicting whether two changes belong together, 
using a training size of $n =$ 320,000 pairs 
(or 800 fine-grained changes), on the \dsOne~dataset. 
Overall, and across all metrics, the
Random Forests algorithm delivers the best results, by a large margin.  The
high {\sc rec} measurement of the \ct{binlogreg} result can be justified by
its equally low {\sc prec}; the classifier marks most of the file changes as
belonging in the same cluster, but few of those decisions are correct.


\begin{table}[ht]
\begin{center}
\caption{Classification performance on \dsOne~by approach}
\label{tab:classifierperf}
\begin{tabular}{l | cc cccc }
{\bf Classifier} & {\sc auc} & {\sc acc} & {\sc prec} & {\sc rec} & {\sc f.measure} & \sc{g.mean} \\ \midrule
 
\rowcolor[gray]{0.97}\ct{binlogreg}  & 0.92 & 0.68 & 0.43 & \bf{0.96} & 0.60 & 0.76\\
\ct{naivebayes} & 0.88 & 0.65 & 0.41 & 0.94 & 0.57 & 0.73\\
\rowcolor[gray]{0.97}\ct{ranforest}  & \bf{0.99} & \bf{0.96} & \bf{0.96} & 0.88 & \bf{0.92} & \bf{0.93} \\ \midrule
\ct{ranforest-trimmed}  & 0.98 & 0.95 & 0.96 & 0.82 & 0.88 & 0.90\\

\end{tabular}
\end{center}
\end{table}

\begin{figure}
  \begin{center}
    \includegraphics[width=\columnwidth]{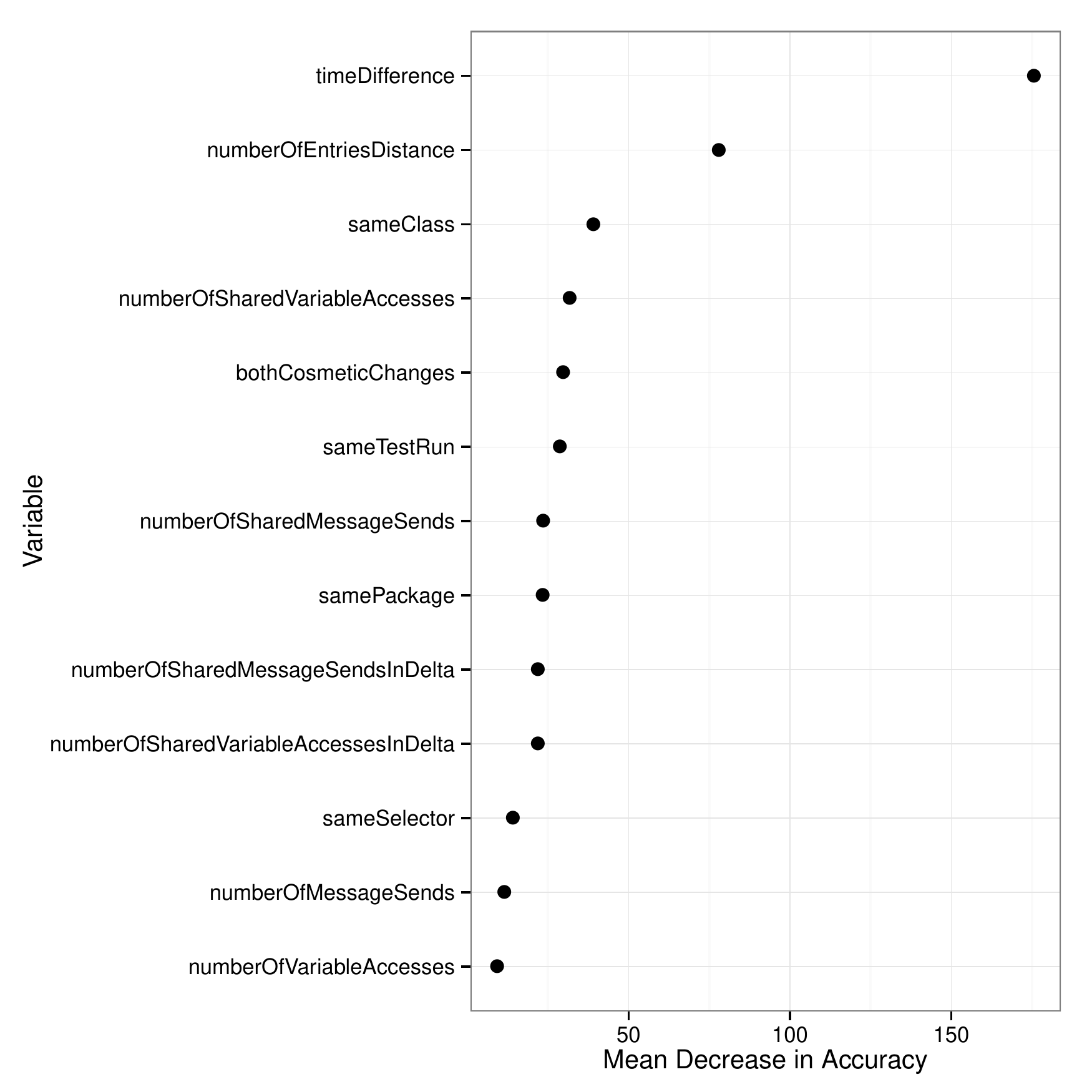}
  \end{center}
  \caption{Voter importance for the random forest classifier.}
  \label{fig:classifperf}
\end{figure}

Once we established that \ct{ranforest} delivers the best results,
we assessed the importance of each voter for its classification result.
We used the process suggested by Genuer \etal~\cite{Genue10}. Specifically, we
run the algorithm 50 times on a randomly selected sample of $10^6$ change pairs,
using a large number of generated trees (500) and trying 5 random variables per split.
Then, we used the mean across 50 runs of the Mean Decrease in Accuracy
metric, as reported by the R implementation of the \textsl{ranforest} algorithm,
to evaluate the importance of each feature. The results can be seen in
Figure~\ref{fig:classifperf}. The three most important voters are:
\begin{inparaenum}[(1)]
\item the time difference between the changes, 
\item the ordered distance of the changes,
\item and whether the changed code belonged in the same class.
\end{inparaenum}
We cannot make inferences about whether the effect of each voter is positive or
negative to the response class; nevertheless, we believe that the results are
indicative of the task-based nature of software development. 

\subsection{How Effective Is Random Forests with the Dominant Voters?}

We answer our second research question by using only the three most
important voters to train the prediction model. The prediction results
are reported in \Cref{tab:classifierperf}, marked as \ct{ranforest-trimmed}. 
We see that even with just those voters we obtain very good prediction results:
The new model is within 3\% of the performance of the model
trained in all metrics.

\begin{figure}[h]
  \begin{center}
    \includegraphics[width=\columnwidth]{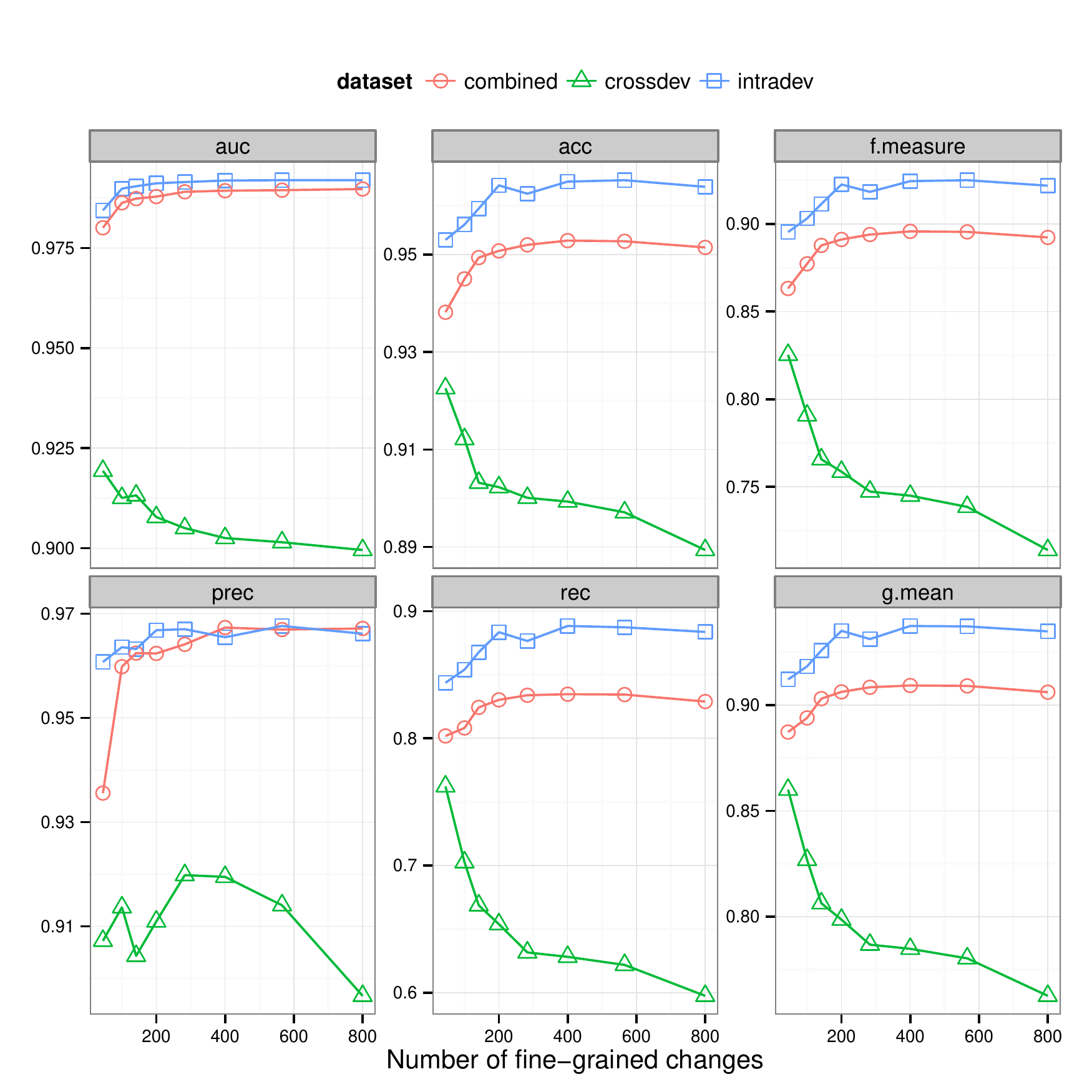}
  \end{center}
  \caption{Dataset performance metrics }
  \label{fig:developers}
\end{figure}

Furthermore, we analyze the impact of the developer who made the changes
on training and testing. We expect that behaviors of developers might be different
and have a significant impact on the model.

We start showing that we obtain the best results when we train and test
from data generated by the same developer (the \ct{intradev} dataset
in \Cref{fig:developers}). This confirms our hypothesis that the behavior
of the specific developer has an impact on the model and the results.
Furthermore, we see that the results are not equally good when training
with data from one developer and testing on the other (the \ct{crossdev}
dataset); moreover we see that as we increase the training size, there is
a drop in performance. This can be attributed to overfitting the model
to the working habits of each individual developer.
Finally, we see that we can train accurate models by combining data 
from multiple developers. In the \ct{combined} dataset, we combine
the data generated by both developers and use this to train the model;
this means that training and testing data
is taken from both samples. \Cref{fig:developers} shows that this dataset
reaches high and stable results; and overfitting seems not present.

What is interesting to note is that the number of fine-grained changes required
for training in both the \textsl{combined} and \textsl{intradev} cases is low:
with 200 changes we can obtain prediction results only 2\% worse (in terms of
\textsf{acc}) on average than if we train with 800 changes. As 200 fine-grained
changes are the equivalent of a few days of work,\footnote{From the data
we recorded, 200 fine-grained code changes correspond to two to five days of work,
depending on the developer's style and pace.} 
we have encouraging evidence that an accurate model can be trained fast 
and deliver good results for a single developer.
Moreover, a pre-trained model with data from multiple developers
might be enough as a starting point for an untangling tool, which could then be
trained to a particular developer's working habits.

Overall, the results show that using the random forest algorithm,
a randomized set of about 200 fine grained changes and a few easy-to-calculate voters, we can train a prediction model that can identify
whether two changes belong in the same commit with an accuracy of 95\%
for a single developer.

\subsection{How Effective Is \approach for Developers?}

We answer research question three by deploying \approach
with developers and recording whether the clustering that it proposes
corresponds to participants' expectations. The dataset \dsTwo, resulting
from this evaluation is described in \Cref{tab:dsTwo}.
We notice that not all the developers coded full time
during the two weeks, thus some produced fewer changes.


\begin{table}[ht]
\begin{center}
\caption{Descriptive statistics of dataset \dsTwo}
\label{tab:dsTwo}
\begin{tabular}{l | rr | rrrr }
\multirow{2}{*}{\bf P\_ID} & \multicolumn{2}{c}{\bf Total number of }& \multicolumn{4}{c}{\bf Changes per cluster}\\
 & {\bf changes} & {\bf clusters} & {\bf Mean} & {\bf Median} & {\bf St. Dev.} & {\bf Max} \\ \midrule
 
 \tommaso & 350 & 22 & 15.9 & 11 & 13.5 & 42\\
\rowcolor[gray]{0.97} \sebastian & 826 & 28 & 29.5 & 3.5 & 50.9 & 228\\
 \anne & 200 & 13 & 15.4 & 10 & 17.3 & 65\\
\rowcolor[gray]{0.97} \vincent & 166 & 12 & 13.8 & 6.5 & 15.6 & 47\\
 \christophe & 347 & 18 & 19.3 & 7 & 27.8 & 88\\
\rowcolor[gray]{0.97} \guillaume & 162 & 11 & 14.7 & 10 & 12.7 & 37\\

\end{tabular}
\end{center}
\end{table}

\begin{figure}[h]
  \begin{center}
    \includegraphics[scale=0.4]{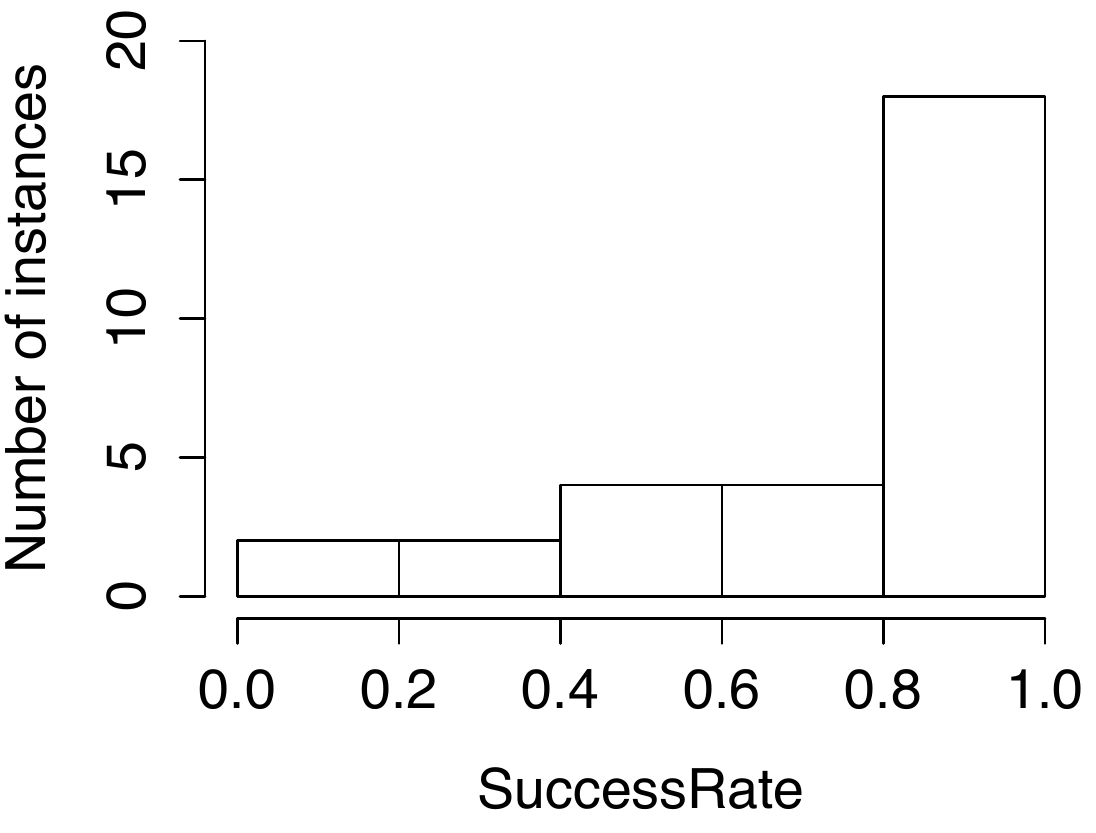}
  \end{center}
  \caption{Frequency of success rate of \approach clustering approach}
  \label{fig:devEvaluation}
\end{figure}

We compared each cluster we proposed to the cluster that
the participant eventually judged as correct to be committed.
The histogram in \Cref{fig:devEvaluation} shows the frequency of the obtained results:
We observed a median\footnote{The results are not normally
distributed, thus we report the median value.} success rate of $0.915$ and an average
of $0.753$ with a standard deviation of $0.30$. By inspecting the instances with a success rate in the range [0,0.4] we could not pinpoint any systematic error; we plan to further address these cases in future work.

We asked developers their opinion on the tool and received diverse feedback.
Most developers were positive [\tommaso, \sebastian, \vincent, \guillaume], 
\eg \tommaso~expressed the feeling that
\quotation{\emph{\approach} guesses correctly the clusters of changes,
also in a big commit were I had 10 different clusters,} and \sebastian~said: \quotation{It works good in many cases, especially for not so big change sets.}
At the same time, most developers [\sebastian, \anne, \vincent, \guillaume] expressed 
concerns with the large amount of fine-grained
information to be processed; they explained that it adds too much noise 
to see not only the last state of a method 
but also all the intermediate modifications to it, especially when belonging
to the same cluster. In the words of \guillaume: \quotation{It was a bit painful to check everything.}

Some participants suggested improvements to the user interface:
For example, \christophe~said that he \quotation{would like to option to delete tasks in the UI},
and \vincent~said: \quotation{I would like to type a name for each task in the UI, as a reminder while I cluster.}




\section{Discussion}

In this section we discuss our results and show how we
mitigate the threats that endanger them.

\subsection{Results}

In the first research question, we asked which voters, or features
are significant to untangle fine-grained code changes.
Despite the fact that we implemented voters along six dimensions, only two dimensions
were significant and contributed to most of the outcome: \emph{code structure}
and \emph{spread}. In particular, the latter dimension has the greatest impact,
by a large margin; the only significant voter in the former dimension 
measured whether the two changes were happening in the same class.
This implies that these voters can be applied 
to other object-oriented programming languages 
regardless of whether they use types or not.
This is a ripe opportunity for testing the approach with different
languages and in different settings. Moreover, although we have
no information about the significance of the voters implemented
by Herzig \etal~\cite{Herz11a,Herz13a}, studies can be designed 
and carried out to determine if and how untangling effectiveness
increases as a result of combining their voters with our significant  ones.

We were surprised by the low impact of many of the voters in
the untangling task: We expected message sending and variable access,
as well as testing information, to contribute more. Since our
initial data analysis was conducted with changes collected by only two
developers, a further study with a larger set of developers for 
generating training data would be useful to confirm or alter this result.

In the second research question, we asked how effective is the best 
performing machine learning algorithm (\ie random forests) when used
with the most significant voters. The results were overall very good.
Expectedly, we achieved the best results when training and testing 
on data from the same developer,
nevertheless cross-developers results are promising and
merged-developers results do not show overfitting; in addition, approximately 200
fine-grained code changes were enough to reach most of the effectiveness.
This implies that training on more developers is necessary to achieve a more general
approach, and there seem to be no risk of overfitting by doing it.
Moreover, ideally every user should
train the approach on her own programming behavior; this seems reasonable
since the training is effective with as little as a few days of work.

In the third research question, we investigate the effectiveness of the whole
approach when deployed to programmers. Considering that the recruited
participants were not used for the training phase, the results are in line with
the effectiveness measured for RQ2. One of the most recurring complaints
was about the large number of changes to be verified and sorted at every commit.
This is due to the fact that we showed all the fine-grained changes recorded,
thus also intermediate states for the same method (when the developer
saved multiple times). We expect this information overload problem to be
mitigated once the approach is stable enough to work correctly in most cases.
Nevertheless, we see a good opportunity for further investigating how
certain fine-grained code changes can be omitted, without losing 
relevant information that would lead to the incompleteness discussed
in \Cref{sec:exSolution}. Moreover, valuable comments were provided about 
the UI of \approach. The UI evaluation goes beyond 
the scope of this paper, but improving the UI can help to have 
an impact on reducing the information overload of fine-grained code changes.



\subsection{Threats to Validity}

\noindent\textbf{Internal Validity.}
Our models and feature selection process are based on a dataset
generated through the actions of two developers. While we have combined the
actions of the developers and shown that they provide very good prediction
performance and the evaluation of the \approach has been overwhelmingly
positive, it is possible that our findings are biased towards the two
developers' working habits.

Bias with respect to developer working habits might also occur in our selection
of evaluation subjects. To reduce this risk, we selected diverse developers, all
of them working in different projects and even in different physical locations.
Thus, we believe the participants represent a heterogeneous enough population
of Pharo developers.

\noindent\textbf{Construct Validity.} 
The notion of task is ambiguous. In particular, each participant can interpret the task granularity differently. For example, consider a single bug fix which is intended to fix two broken features. The participant could consider the changes either as two individual tasks, or everything as a single bug-fixing task. For mitigating this risk, we prepared a screencast with an example for users trying to establish a common criterion for task granularity. Moreover, we kept in close contact with users for answering any doubt. However, this ambiguity in the definition of task does not reduce the precision of our success metric for answering RQ3 (\emph{SuccessRate}), since it represents each user expectation: it compares \approach's clustering with the participant's expected clustering.

The clustering computed by \approach may have influenced participants. When users had to evaluate the computed clustering (as shown in \Cref{screenshot-training}), the initial clustering might have biased their answers.

\noindent\textbf{External Validity.} 
We used a specific platform (Pharo) and language environment (Smalltalk) to
facilitate our study. A specific language may dictate a specific working style.
For example, in a typed language setting, an IDE would immediately mark as
erroneous cases where a type signature has changed and not all uses have been
adapted, therefore prompting the developer to fix such cases. Therefore our
results may not be generalizable to all languages or working environments.



\section{Related Works}

The impact of tangled changes has been reported in several contexts:
The inspiring work by Herzig \etal~\cite{Herz13a}, reported that
at least 16.5\% of all source files in the datasets they considered were
incorrectly associated with bug reports when ignoring the existence of 
tangled change sets. In a large-scale study done at Microsoft on 
how developers understand code changes, Tao \etal reported that 
developers find it important for understanding to decompose changes into the
individual development issues, but there is currently no tool support for doing so~\cite{Tao2012a}.
Bacchelli and Bird reported that tangled changes in code to be reviewed
often cause low quality reviews or require longer time to review~\cite{Bacc2013a}.

Herzig \etal were the first to implement an algorithm to automatically generate untangled commits given a tangled one. Their work greatly inspired our research. However, we see some limitations to their work that we explained in \Cref{sec:problem}: static-analysis dependency, incompleteness, and artificiality. The main differences with our work is that: 
\begin{inparaenum}[(1)]
\item we count with fine-grained timing information of code changes as well as IDE events like test runs; 
\item we work in a dynamically-typed language; 
\item we evaluated our approach with developers.
\end{inparaenum}

Another source of inspiration comes from Robbes, who created
a fine-grained change model of software evolution based on three
principles \cite{Robb08b}:
\begin{inparaenum}[(1)]
\item a program state needs to be represented accurately by an
  Abstract Syntax Tree (AST);
\item a program's history is a sequence of changes, each one producing
  a program state (an AST) and changes can be composed into
  higher-level changes;
\item changes should be recorded by the IDE as they happen, not
  recovered from a VCS.
\end{inparaenum} Robbes \etal show how a fine-grained change
model can better detect \emph{logical coupling} between
classes \cite{Robb08a}. Their article presents new measures of
logical coupling that we consider as a future extension of our voters.

Steinert \etal propose CoExist, an approach and associated tool set to
navigate the different states of a project based on its fine-grained
changes \cite{Stein12b}.
CoExist's tool suite allows for
reverting any fined-grained change at the project level, comparing
different states of a program, localizing the cause of a failing test
in the development history, and reassembling changes to share untangled commits.
Automatic clustering of dependent
fine-grained changes to create untangled commits is left as future
work. Our work can be seen as an extension of CoExist tool suite in
this direction, despite its totally unrelated implementation.

Wloka \etal presented a program analysis technique to identify
committable changes that can be released early, without causing
failures of existing tests \cite{Wlok09a}. Wloka remarks that an
untangling algorithm would clearly benefit from having a model with a
more accurate concept of change to add context information for
individual change operations. Beyond our \ct{Same Test Run} voter, we leverage more the results of unit-test execution to cluster
related changes.




\section{Conclusion}


In this paper, we have devised and evaluated \approach, an approach whose ultimate
goal is to help developers share self-contained changes that are well-decomposed
into individual tasks. 
We build on the shoulders of others, and expand previous work by:
\begin{inparaenum}[(1)]
\item Working in an untyped language setting where static code analyses are more limited;
\item considering fine-grained code change information gathered during development; and
\item evaluating the resulting approach both on data generated by programmers who manually labeled it and with programmers working on real development tasks.
\end{inparaenum}

Our results show that three features are especially important to perform clustering of fine-grained code changes: the time between the changes, the number of other modifications between the changes, and whether the changes modify the same class. By testing the features on historical data manually labeled by developers, we obtained good results (over 88\% of accuracy in the worst case) in determining whether two changes should be together. When deploying our approach with new developers, we obtained a median success rate of 91\%.

Overall, this paper makes the following main contributions:

\begin{enumerate}

\item An analysis of the current points for improvement in the state of the art in untangling code changes.

\item A publicly available\footnote{Available at: \url{http://dx.doi.org/10.6084/m9.figshare.1241571}} dataset of fine-grained code changes collected by recording the development sessions of two developers over the course of four months, and the corresponding manual clustering.

\item The creation of different features/voters and their evaluation, based on the aforementioned dataset, using machine learning approaches to model and classify pairs of fine-grained code changes, resulting in good accuracy results.

\item The creation of an approach, \approach, and corresponding tool implementation, \tool,\footnote{Available at: \url{http://smalltalkhub.com/\#!/~MartinDias/EpiceaTaskClusterer}} to untangle fine-grained code changes into clusters based on the three best voters and the best performing machine learning algorithm.

\item The deployment and a two-week evaluation of \approach with developers with good results.

\end{enumerate}


\section*{Acknowledgements} We thank our study participants for their feedback and the European Smalltalk User Group\footnote{ESUG: \url{http://esug.org}} for its support.

\bibliographystyle{IEEEtran}
\bibliography{changeuntangling}

\end{document}